# Pushing the limits of ab-initio-NEGF transport using efficient dissipative Mode-Space algorithms for realistic simulations of low-dimensional semiconductors including their oxide interfaces

Aryan Afzalian, Fabian Ducry
imec, Leuven, Belgium, aryan.afzalian@imec.be

*Abstract*—We investigate the trade-offs between accuracy and efficiency for several flavors of the dissipative mode-space NEGF algorithm with the self-consistent Born approximation for DFT Hamiltonians. Using these models, we then demonstrate the dissipative self-consistent DFT-NEGF simulations of realistic 2D-material devices including their oxide interfaces with large slab dimensions (up to 1000 atoms) and up to several 100,000 atoms in the full device, pushing the limit of ab-initio transport.

## Introduction

Due to scaling, the typical MOSFET transistor dimensions have shrunk deep in the nm-range regime and novel low-dimensional (LD) materials such as 2D transition-metal dichalcogenides (TMDc) or ultra-thin Si or Ge nanosheets are explored [1,2,3,4]. Owing to their ultra-thin nature, interfaces with e.g., metal contacts and gate oxides (Fig. 1a), usually dominate their transport characteristics rather than intrinsic LD material properties.

Ab-initio atomistic methods, such as Density-Functional-Theory (DFT) modeling, are crucial to understand the physics, practical challenges, and proof-of-concept of devices made of novel materials and their interfaces. Particularly, because there is typically a lack of experimental data and mature processing capability. These techniques are, however, computationally costly, especially when having to tackle large supercells resulting of the interface between the LD-materials and their crystalline or amorphous contact or oxide (Fig. 1).

Mode-Space (MS) methods have demonstrated their capability to significantly speed-up, not only effective mass, but also atomistic computations with tight-binding or even DFT Hamiltonians (H) [3,4,5]. One point that requires special care in the MS algorithm is the treatment of electron-phonon (e-ph) scattering, as local scattering mechanisms in real space become, in principle, non-local in MS due to mode-coupling [6,7].

In the past, we have observed that this mode-coupling is a second order effect that can usually be neglected for computing the scattering self-energy (but not for the carrier or current densities calculations) of localised bases, such as effective masses or Slater-Koster nearest-neighbour tight-binding H [6,7]. The situation is less clear with DFT bases, as the wave function is typically delocalized, and long-range coupling is present. In our quantum transport solver ATOMOS [1,2], we have developed a careful, efficient, and memory-lean implementation of the atomistic mode-space (MS) NEGF algorithm that works both for orthogonal- [5] and non-orthogonal-H [4] models. We included various e-ph scattering flavors, using the self-consistent Born formalism, that neglect or include mode-coupling. We applied them to various materials, using DFT-based H, in order to benchmark these methods against RS simulations. Finally, using the MS model, we demonstrate the self-consistent DFT-NEGF simulations of very large devices including e-ph scattering, pushing the limit of dissipative ab-initio transport.

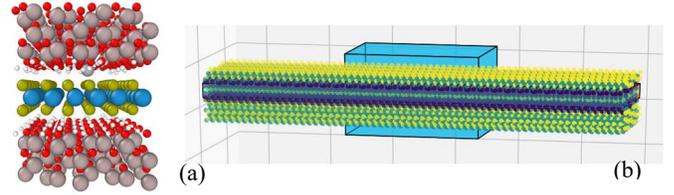

Fig. 1. a) Atomistic view of a DFT-relaxed monolayer $WS_2$ slab including a crystalline $Al_2O_3$ oxide interface with >700 atoms. b) NIN $WS_2$ double-gate (DG) NS MOSFET made of the $WS_2$ slabs including a crystalline $Al_2O_3$ oxide interface simulated with ATOMOS.

## Methods

To compute the DFT-based H, in this work, we used the projector-augmented-wave (PAW) DFT, followed by a Wannierization step, with the QUANTUM ESPRESSO package [8] and the generalized gradient approximation (GGA) with the optB86b exchange-correlation (XC) functionals [9] for the free standing $WS_2$ simulations as fully described in [1]. We also used the linear-combination-of-atomic-orbitals (LCAO) DFT with a non-orthogonal model (as described in [2]) and the CP2K Package [10], using GGA and PBE XC for $WS_2$-$Al_2O_3$ or meta-GGA and the PBE0-TC-LRC [11] hybrid functional with 25% of Hartree-Fock exchange and 75% PBE for Si and Ge in order to get a correct bandgap.

Concerning MS-NEGF with scattering, we use capital letters for RS quantities and small letters for MS ones. Within the self-consistent Born approximation, assuming the phonons stay in equilibrium, and that the e-ph coupling matrix $M_q$ is expressed in an orthogonal basis (e.g., by a deformation potential), the RS self-energy for the electron-phonon interaction is [2, 6]:

$$\Sigma^<_{S,\perp}(r_i, r_j, E) = \int \frac{dq}{(2\pi)^3} e^{iq.(r_i-r_j)} |M_q|^2 \\ \times (N_q + \tfrac{1}{2} \pm \tfrac{1}{2}) G^<_\perp(r_i, r_j, E \pm \hbar\omega_q) \quad (1)$$

$$G_\perp^<(E) = G(E) \times S \quad (2)$$

where $q$ and $\omega_q$ are the phonon wave vector and angular frequency, $\hbar$ is the reduced Plank's constant, $N_q$ is the phonon occupation number. $S$ is the overlap matrix, that is equal to the unity matrix in an orthogonal model. In the case of a non-orthogonal model, the scattering self-energy in the non-orthogonal basis can be obtained from the orthogonal one, e.g., in a symmetric fashion [2]:

$$\Sigma_S^< = \tfrac{1}{2}\left(S\Sigma_{S,\perp}^< + \Sigma_{S,\perp}^< S\right) \quad (3)$$

Assuming acoustic elastic phonons and inelastic optical phonons modelled by deformation potentials, under the usual approximations, these scattering mechanisms are local in RS [6]. In MS, to compute $\sigma_S^<$, it is, hence, possible to up-convert $g^<$ to RS using,

$$G^< = U g^< U^\dagger \quad (4)$$

then compute $\Sigma_S^<$ using eq. (1), (2) and (3), and, finally, compute $\sigma_S^<$ using the inverse transform:

$$\sigma_S^< = U^\dagger \Sigma_S^< U \quad (5)$$

In eq. (4) and (5), $U$ of size $N \times n_{MS}$ is a pre-optimized block-diagonal unitary transformation matrix to change from the original real space of size $N$ to the reduced mode space of size $n_{MS}$ ($n_{MS} < N$) [3,4,5].

In this method (here MS-UPDOWN) the speed of NEGF is typically dominated by the up- and down-conversions from MS to RS. We therefore use a dedicated pool of workers to perform these operations in parallel in our implementation. Still, for efficiency, a further simplification is required, namely neglecting the mode coupling in the down-conversion, i.e., only the local (diagonal) terms of $\sigma_S^<$ are considered in eq. (5).

Concerning the literature, however, an equivalent and faster direct MS expression based on a form factor ($F$) method (MS-FF) is used [6]:

$$\sigma_{S,\perp,mn}^<(x_i,x_j,E) = \int \frac{dq}{(2\pi)^3} e^{iq_x(x_i-x_j)} |M_q|^2$$
$$\times \sum_{kl}(N_q + \tfrac{1}{2} \pm \tfrac{1}{2}) g_{\perp,kl}^<(x_i,x_j,E \pm \omega q) \times F_{mn}^{kl}(x_i,x_j,q_t) \quad (6)$$

It is typically employed in a simpler form, i.e., neglecting both up- and down-conversion mode coupling (i.e., conserving only terms for $m = n$ and $k = l$) as, including the non-local scattering in $F$ quadratically scales with the number of modes $n_{MS}$ and is prohibitively expensive for large atomistic simulations [7]. Using this local approximation, the scattering self-energy $\sigma_S^<$ is diagonal and the form factor $F$ simplifies to [6,7]:

$$F_{mm}^{kk}(x_i) = \int \Psi_m^*(y,z;x_i)\, \Psi_m(y,z;x_i)$$
$$\times \Psi_k(y,z;x_i)\Psi_k^*(y,z;x_i) dy dz \quad (7)$$

where $\Psi_i$ is the i$^{th}$ eigenvector (i$^{th}$ column) in the corresponding diagonal block of the MS transformation basis $U$. Again, in the case of a non-orthogonal model, the scattering self-energy in the non-orthogonal basis can be obtained from the orthogonal one, using the MS equivalent of eq. (3).

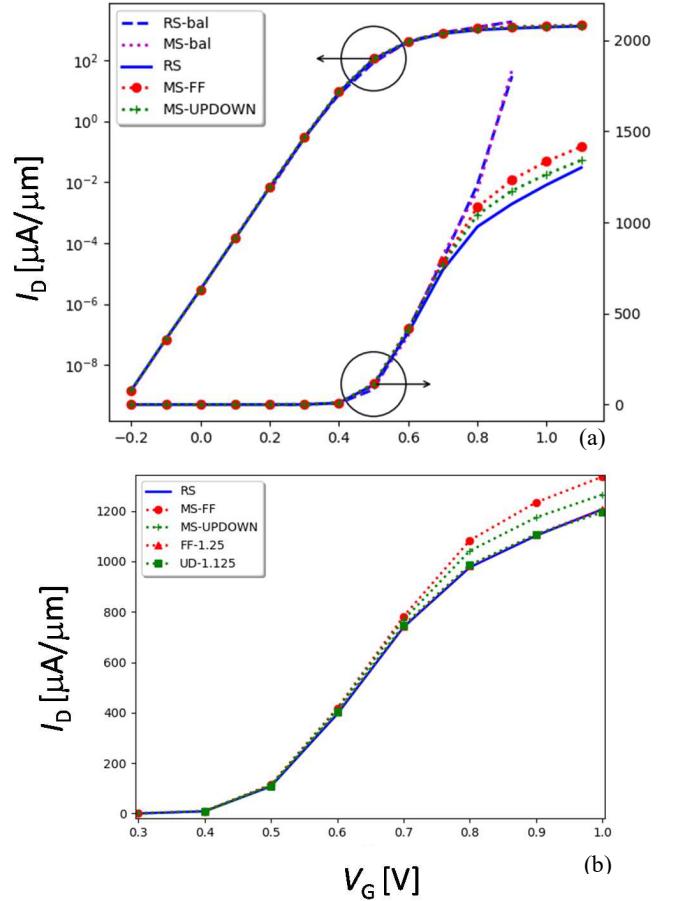

Fig. 2. **Monolayer-WS$_2$ DG NS**: DFT-NEGF-simulated drain current – gate voltage $I_D(V_G)$ characteristics for a WS$_2$ NS nMOSFET. a) $I_D(V_G)$ in logarithmic and linear scale with RS and MS NEGF using ballistic (bal) or e-ph scattering methods (for MS using the MS-FF or MS-UPDOWN with $\eta$ = 1). b) Zoom on the $I_D(V_G)$ in linear scale with RS and MS NEGF with e-ph scattering, for MS using the MS-FF or MS-UPDOWN with $\eta$ = 1 or the RS-matched MS-FF with $\eta$ = 1.25 (FF-1.25) or MS-UPDOWN with $\eta$ = 1.125 (UD-1.125). $L$ = 10 nm. $V_D$ = 0.6 V. The current is normalized by the gate perimeter (i.e, the current per width and per gate is plotted here).

### Results

DFT-based NEGF simulations of Nanosheet (NS) transistors made of WS$_2$, with and without Al$_2$O$_3$, as well as Si and Ge were performed using ATOMOS with the RS and MS methods both ballistically (bal) and including scattering (e-ph), for MS using the various flavours implemented. Similar results were observed in all cases. Large speed up were obtained using the MS models (from ~ 20 to 2000×). In the ballistic case, RS and MS $I_D(V_G)$ characteristics were in good agreement (Fig. 2a).

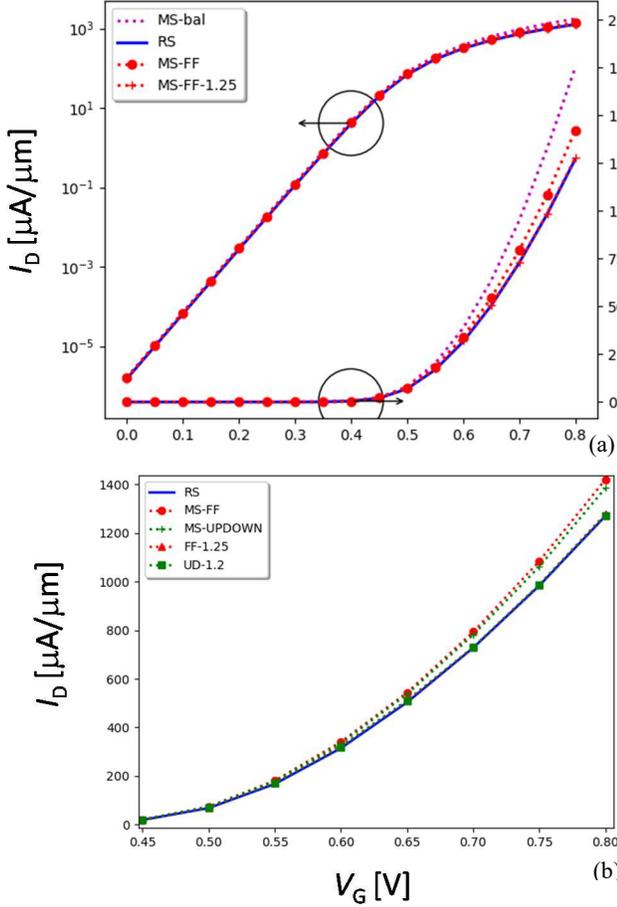

Fig. 3. **3.5-nm-thick [110]-oriented Si DG NS**: DFT-NEGF-simulated $I_D(V_G)$ characteristics for a Si NS nMOSFET. a) $I_D(V_G)$ in logarithmic and linear scale with RS and MS NEGF using ballistic (bal) or e-ph scattering methods (for MS using the MS-FF or UPDOWN with $\eta = 1$). b) Zoom on the $I_D(V_G)$ in linear scale with RS and MS NEGF including e-ph scattering, for MS using the MS-FF or MS-UPDOWN with $\eta = 1$ or the RS-matched MS-FF with $\eta = 1.25$ (FF-1.25) or UPDOWN with $\eta = 1.2$ (UD-1.2). $L = 10$ nm. $V_D = 0.6$ V.

With scattering, an overestimation of the on current by a few percents was typically observed for MS, the effect was more pronounced in the FF-MS method, indicating that neglecting the mode-coupling in a DFT basis underestimates the scattering rate (Fig. 2, 3 and 4). In all MS cases, using a $\eta$ factor to compensate for neglecting the non-local scattering, by increasing the local deformation potential values by 25 to 30% (10 to 20%) in the MS-FF (MS-UPDOWN respectively), was sufficient, as shown on Fig. 2b (with MS-FF and $\eta = 1.25$ or MS-UPDOWN and $\eta = 1.125$) and 3b (with MS-FF and $\eta = 1.25$ or MS-UPDOWN and $\eta = 1.2$) for the WS$_2$ and Si cases, respectively. For Ge, a very similar $\eta$ factor for the MS-FF model is found. In this case, also, the minimum current, related to the band-to-band tunnelling (BTBT) current limit in off-sate is well reproduced (Fig. 4). The required $\eta$ factor seems largely independent of the operating biases or device dimensions.

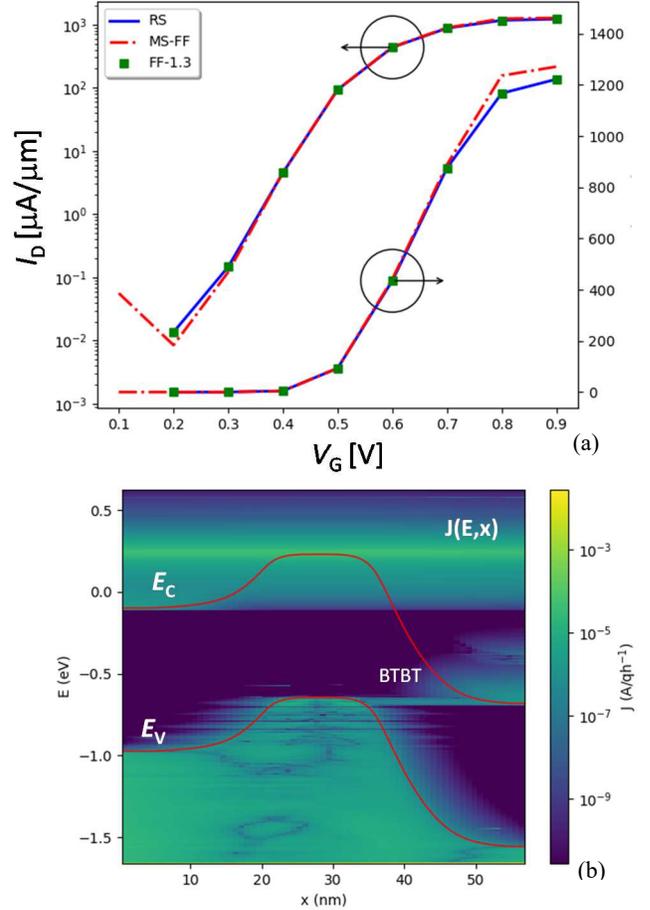

Fig. 4. **3.5-nm-thick [110]-oriented Ge DG NS**: DFT-NEGF-simulated $I_D(V_G)$ characteristics for a Ge NS nMOSFET. a) $I_D(V_G)$ in logarithmic and linear scale with RS and MS NEGF with e-ph scattering methods (for MS using the MS-FF with $\eta = 1$, or the RS-matched MS-FF with $\eta = 1.3$). b) Current spectrum $J(E,x)$ (surface plot), as well as conduction-band ($E_C$) and valence band ($E_V$) edges (-) along the channel direction, $x$, of the Ge NS in off state, at $V_G = 0.2$ V, i.e., close to the BTBT current limit $I_{MIN}$. Both the thermionic current above $E_C$ and the BTBT curent between $E_V$ in the channel and $E_C$ at drain-side can be observed. $L = 16$ nm. $V_D = 0.6$ V.

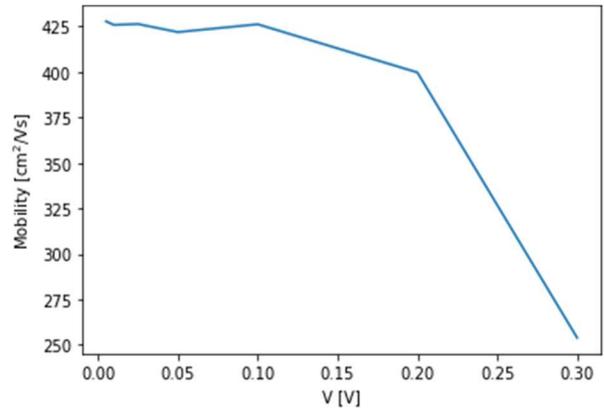

Fig. 5. Apparent mobility vs. applied voltage extracted from a 300-nm-long resistor made with the WS$_2$ - Al$_2$O$_3$ slabs from Fig. 1a. The device is made of more than 100,000 atoms and was simulated using the DFT MS-FF NEGF model with a calibrated $\eta = 1.3$.

Using the calibrated η factors, realistic quantum transport simulations of LD NS devices including their oxide interface with large slab dimensions (up to 1000 atoms) and up to several 100,000 atoms in the full device were performed, using accurate dissipative NEGF methods and a full DFT basis.

Fig. 5 shows the apparent mobility vs. applied voltage extracted from a 300-nm-long $WS_2$ resistor made of $WS_2$-$Al_2O_3$ slabs (Fig. 1a). Fig. 6 shows the $I_D(V_G)$ characteristics for n-type NS MOSFETs made with the 3.5-nm-thick-Si vs. channel length, $L$, ranging from 10 to 200 nm.

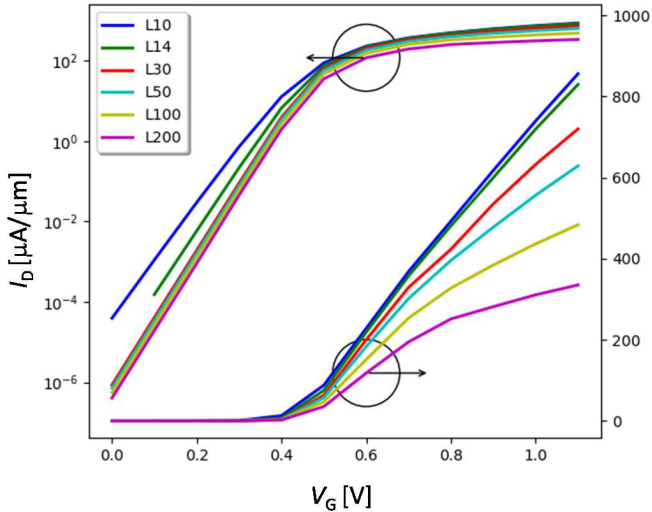

Fig. 6. Dissipative DFT-NEGF-simulated $I_D(V_G)$ characteristics for n-type NS MOSFETs made with the 3.5-nm-thick-Si vs. channel length, $L$, ($L$ ranging from 10 to 200 nm as indicated in the legend) simulated with the calibrated MS-FF NEGF model. $V_D$ = 0.05 V.

## Conclusions

We have developed an efficient implementation of the atomistic MS-NEGF algorithm in our quantum transport solver ATOMOS. We included various e-ph scattering flavors, using the self-consistent Born formalism, that neglect or include mode-coupling for the scattering self-energy computation.

We benchmarked against RS simulations two of these flavors, i.e., the MS-FF, that fully neglects mode coupling, and MS-UPDOWN, that neglects mode coupling in the down-conversion only, using DFT-based H from various materials.

With scattering, an overestimation of the on current by a few percents was typically observed for MS, the effect was more pronounced in the FF-MS method, indicating that neglecting the mode-coupling in a DFT basis underestimates the scattering rate.

In all MS cases, using a small η factor to compensate for neglecting the non-local e-ph scattering, by increasing the local deformation potential values, was sufficient to accurately reproduce the RS results.

The required η factor seems largely independent of the operating biases or device dimensions.

Using the MS model, we demonstrate the self-consistent DFT-NEGF simulations of very large devices including e-ph scattering, pushing the limit of dissipative ab initio transport.